\documentstyle[preprint,eqsecnum,aps,epsf]{revtex}	

\newif\iftightenlines\tightenlinesfalse
\tightenlines\tightenlinestrue

\begin{document}
%
\def\pT{p_T^{\phantom{7}}}
\def\MW{M_W^{\phantom{7}}}
\def\ET{E_T^{\phantom{7}}}
\def\bh{\bar h}
\def\lm{\,{\rm lm}}
\def\lo{\lambda_1}                                              
\def\lt{\lambda_2}
\def\pslt{p\llap/_T}
\def\eslt{E\llap/_T}
\def\to{\rightarrow}
\def\Re{{\cal R \mskip-4mu \lower.1ex \hbox{\it e}}\,}
\def\Im{{\cal I \mskip-5mu \lower.1ex \hbox{\it m}}\,}
\def\SU{SU(2)$\times$U(1)$_Y$}
\def\te{\tilde e}
\def\tl{\tilde l}
\def\ttau{\tilde \tau}
\def\tg{\tilde g}
\def\tga{\tilde \gamma}
\def\tnu{\tilde\nu}
\def\tell{\tilde\ell}
\def\tq{\tilde q}
\def\tt{\tilde t}
\def\tw{\widetilde W}
\def\tz{\widetilde Z}
\def\cmsec{{\rm cm^{-2}s^{-1}}}
\def\sgn{\mathop{\rm sgn}}
\hyphenation{mssm}
\def\ds{\displaystyle}
\def\ts{${\strut\atop\strut}$}
%
\preprint{\vbox{\baselineskip=14pt%
   \rightline{FSU-HEP-960802}\break 
      \rightline{UCD-96-22}
   \rightline{UH-511-853-96}
}}
\title{IMPACT OF HADRONIC DECAYS OF THE LIGHTEST \\ 
NEUTRALINO ON THE REACH OF THE CERN LHC
}
\author{Howard Baer$^1$, Chih-hao Chen$^2$ and Xerxes Tata$^3$}
\address{
$^1$Department of Physics,
Florida State University,
Tallahassee, FL 32306, U.S.A.}
\address{
$^2$ Davis Institute for High Energy Physics, 
University of California, Davis,
CA 95616, U.S.A.}

\address{
$^3$Department of Physics and Astronomy,
University of Hawaii,
Honolulu, HI 96822, U.S.A}
\date{\today}
\maketitle
\begin{abstract}
If $R$-parity is not conserved, the lightest supersymmetric particle (LSP)
could decay via lepton number violating or baryon number violating
interactions. The latter case is particularly insidious since it leads
to a reduction of the $\eslt$ as well as leptonic signals for supersymmetry.
We evaluate cross sections for jets plus $\eslt$, 1$\ell$, $2\ell$ 
(same-sign and opposite sign) and $3\ell$ event topologies that result 
from the simultaneous production of
all sparticles at the CERN Large Hadron Collider (LHC),
assuming that the lightest
supersymmetric particle $\tz_1$ decays hadronically inside the detector via
$R$-parity violating interactions. We assume that these interactions do
not affect the production rates or decays of other sparticles.
We examine the SUSY reach of the LHC for this ``pessimistic'' scenario,
and show that experiments at the LHC will still be able to search for gluinos
and squarks as heavy as 1 TeV, given just 10 fb$^{-1}$ of integrated
luminosity, even with cuts designed to explore the
canonical SUSY framework with a {\it conserved} $R$-parity.
\end{abstract}

\medskip



It is generally accepted that a decisive search for weak scale supersymmetry
(SUSY) will require the direct exploration of the TeV scale. This
would be possible, for instance, at the CERN
Large Hadron Collider (LHC) where experiments should be able to probe
gluino and squark masses up to about 2~TeV \cite{CMS,ATLAS,BCPT}.
The large reach of the LHC provides
a comfortable safety margin beyond the, admittedly subjective, upper limits
of $\sim 800-1000$~GeV on sparticle masses\cite{FINETUN}
from the requirement that SUSY stabilize the scalar
electroweak symmetry breaking sector. These analyses of the reach of the LHC
are, however, performed within the framework of the minimal supersymmetric
model where it is assumed that the three gaugino masses and the
various sfermion masses unify
at some ultra-high energy scale
$M_X \sim M_{GUT}$. Moreover, it is also assumed
that $R$-parity is conserved so that the
lightest supersymmetric particle (taken to be the lightest neutralino, $\tz_1$)
escapes detection, leading to the canonical $\eslt$ signature for SUSY. 
In view of the importance of the issue, 
it is worthwhile
to ask how the LHC reach would be altered if these
assumptions are relaxed: in particular, could it happen that signals from
kinematically accessible sparticles can remain hidden at the LHC?

The gaugino and sfermion mass unification conditions mainly imply that
coloured sparticles are heavier than their colourless counterparts; i.e.
gluinos are heavier than charginos and neutralinos, and squarks are not
lighter than sleptons. If the situation were reversed and the coloured
sparticles are light, they would be even more copiously produced at the
LHC, and so, would be detectable via the $\eslt$ signature as long as
we assume that the lightest SUSY particle (LSP) 
is electrically and colour neutral and escapes experimental detection.
SUSY phenomenology would be quite different since the cascade decays
of gluinos and squarks which lead to the multilepton signatures would be
suppressed, and further, the production of colourless particles ({\it e.g.}
chargino and neutralino production) would dominantly lead to hadronic final
states if the gluino were lighter than $\tw_1$ or $\tz_2$. 
It is, however,
hard to imagine that (barring any artificial degeneracies that drastically
reduce the visible energy in all SUSY events)
gluino and squark signals would escape detection\cite{FN1},
at least in the $\eslt$ channel.

In our study of whether SUSY can possibly remain hidden at the LHC, we
are thus led to consider scenarios where the LSP decays
inside the detector via explicit\cite{FN2} $R$-parity
violating interactions\cite{RPV},
so that $\eslt$, the usual hallmark of SUSY events, is greatly degraded.
The phenomenology of $R$-parity violating models can be very different
from that of the minimal model. $R$-violating interactions, if they are
sufficiently strong, can alter the decay patterns of all sparticles. These
interactions also lead to new sparticle production mechanisms: in particular,
squarks and sleptons can be singly produced as resonances at
colliders.
The resulting modifications are, in general, sensitively dependent on the strength and
flavour structure of the $R$-violating interactions, but are negligible
if the corresponding couplings are much smaller than gauge couplings. We  
will assume this to be the case in our analysis. Then, 
the only impact of these $R$-parity violating interactions on collider
phenomenology is that they cause the LSP to decay inside the detector, thereby
greatly reducing the $\eslt$ signal. 

With the field content of the minimal model, $R$-parity violation can
occur via renormalizable operators that also violate either lepton or
baryon number conservation. While some of these couplings are
indeed significantly constrained by experiment\cite{RPVCONST}
(non-observation of proton 
decay strongly constrains the simultaneous presence of baryon and lepton
number violating operators),
it is possible to build perfectly viable $R$-violating 
models, especially if we assume that
these $R$-violating couplings are much smaller than gauge couplings (but large
enough so that the LSP decays within the detector). The decay patterns
of the LSP depend on the structure of $R$-violating interactions. If
$R$-parity conservation is broken by lepton number violating operators, the
LSP decays either via $\tz_1 \to \ell_i\bar{\ell_j}\nu$ or via $\tz_1 \to \ell
q_i\bar{q_j},
\nu q_i \bar{q_j}$ ($i,j$ denote family indices)
with parameter-dependent branching fractions. The main point
to note is that the decays of each LSP lead to additional isolated leptons
in the final state. In the favourable case where the LSP only decays
into electrons and muons (say via $\tz_1 \to \bar{e}\mu\nu_e$, for example), 
every
SUSY event contains at least four leptons from the decays of the two LSPs,
and can readily be separated from Standard Model backgrounds. In this case,
experiments at even the 
Tevatron Main
Injector (MI) upgrade at Fermilab should be (indirectly) sensitive to 
gluinos as heavy as 700-800~GeV \cite{BKT}, 
and the reach of the LHC should be truly enormous.
If, on the other hand, $R$-parity is violated by
baryon number violating operators, the LSP can only decay purely hadronically
via $\tz_1 \to q_i q_j q_k$ and its charge conjugate mode. 
The hadronic decays of the LSP impact upon
the SUSY signal in two distinct ways. First, the usual $\eslt$ signal is 
greatly reduced as we have already mentioned. Second, the additional 
hadronic activity from the decays of the LSP make it more difficult for
the leptons from the cascade decays of gluinos and squarks to remain
isolated so that the SUSY reach via leptonic channels is also reduced.
Gluinos heavier than about 200 GeV could evade detection at the MI\cite{BKT}
if this scenario is operative. 

The purpose of this paper is to examine the
reach of the LHC within the same scenario, with a view to see whether
sparticle signals could possibly be missed in LHC experiments\cite{EARLY}.
It seems reasonable to suppose that the first searches for SUSY at the
LHC will be performed with the minimal ($R$-parity conserving) framework
in mind; {\it i.e.} using cuts roughly along the lines of the analyses
in Ref.\cite{CMS,ATLAS,BCPT} which gave similar results for the SUSY
mass reach within the minimal model framework.
Here, we therefore examine the SUSY reach of the LHC
in the $\eslt$, $1\ell$, opposite-sign (OS) dilepton, same-sign (SS) dilepton
and $3\ell$ channels,
using the same cuts (we do not list these for brevity)
as in our earlier analyses \cite{BCPT} of the $R$-parity conserving case. 
We stress that these cuts are not
optimized for searching for SUSY when $R$-parity is violated. Our
purpose here is to roughly delineate the region of parameter space where the
initial searches for supersymmetry will lead to an observable signal. 

For definiteness (and to be able to compare with our previous analyses
within the $R$-parity conserving framework), we use the minimal SUGRA model
as discussed in
Ref.\cite{BCPT}, and assume that the superpotential
term $\lambda''_{cds}C^cD^cS^c$ is the only
source of $R$-parity violation\cite{BKT}.
We note that as long as the coupling $\lambda''_{cds}$ is much smaller
than the gauge couplings, it makes a negligible effect on the masses and
couplings obtained using renormalization group evolution, so that the
production mechanisms and decay patterns of all sparticles (other than the LSP)
are unaltered
from minimal model expectations. 
The LSP, however, decays via $\tz_1 \to cds$ or $\tz_1 \to 
\bar{c}\bar{d}\bar{s}$, with the two modes having the same branching ratio
by CP invariance. Our results are insensitive to the assumed flavour
structure of the $R$-violating
interaction as long as we do not attempt any flavour tagging, and should
give the maximum degradation of the signal, at least as long as
these interactions do not affect the decays of heavier superparticles.
The model is then completely specified
by four parameters which we take to be the common scalar mass $m_0$,
a common gaugino mass $m_{1/2}$, $\tan\beta$ and $A_0$ together with
the sign of $\mu$. The magnitude of the $R$-violating coupling only
affects the LSP lifetime which is irrelevant as long as it decays 
inside the detector\cite{FN3}. 

Following Ref.\cite{BCPT}, for a 100~GeV $\times$ 100~GeV
grid of points in the $m_0-m_{1/2}$ plane,
we have used ISAJET 7.20 \cite{ISAJET} to simultaneously
generate all $2 \to 2$ sparticle subprocesses and the subsequent cascade decay
chains as given by the model. We use CTEQ2L structure functions in our
computations of the cross sections\cite{CTEQ}.
The $3q$ decays of the LSPs produced at the
end of the cascade are implemented by explicit addition of decay modes
to the ISAJET decay table. We use
the toy calorimeter simulation package ISAPLT with the same hadronic and
electromagnetic resolution smearing as in our previous analysis. Events are
classified by their isolated lepton ($e$ and $\mu$) 
content into the $\eslt$, $1\ell$, OS, SS and
$3\ell$ topologies using exactly the same cuts as before \cite{BCPT}. Standard
Model physics background levels
from $t\bar{t}$, $W$ + jet, $Z$ + jet production,
QCD and vector boson pair production have been shown
in Ref.\cite{BCPT} and will not be reproduced here. We consider a signal
to be observable if  (for any value of the floating cut parameter $E_T^c$
defined in Ref.\cite{BCPT})
{\it i})$N_S >5\sqrt{N_B}$, {\it ii}) $N_S>0.2N_B$
and {\it iii}) there are at least 5 signal events in 10~$fb^{-1}$ of
integrated luminosity at the LHC. Here, $N_S$ is the surviving number of 
signal events and $N_B$ is the surviving number of background events for
our choice of integrated luminosity.

The region of the $m_0-m_{1/2}$ plane where there is an observable signal
in the various leptonic channels is shown by the
various contours in Fig.~1 for {\it a}) $\tan\beta=2, \mu < 0$, {\it b})
$\tan\beta=2, \mu > 0$, {\it c}) $\tan\beta=10, \mu < 0$, and
{\it d})$\tan\beta=10, \mu > 0$. We have taken $A_0 = 0$ and fixed
$m_t=170$~GeV. The bricked regions are excluded by theoretical constaints, 
and the shaded regions are excluded by experimental constraints, as 
discussed in Ref. \cite{BCPT}-- the only difference is that we have updated 
the chargino mass limit to $m_{\tw_1}>65$ GeV in accord with recent LEP2 
data\cite{LEP2}.
Various sparticle mass contours are shown in Ref. \cite{BCPT}; here, we
show only the $m_{\tg}=1000$ GeV and $m_{\tq}=1000$ GeV contours for clarity.
Several comments are worth noting:
\begin{itemize}
\item The outer envelope of the maximal reach contours in all four frames
extends beyond the $m_{\tg}$ and $m_{\tq}=1000$ GeV contours.
This implies that
squarks and gluinos with masses smaller than 1~TeV should be observable
via the canonical leptonic search channels at the LHC even in 
this ``pessimistic'' scenario.
If $m_0 \leq 300-400$~GeV, the LHC reach extends out to $m_{1/2} \sim
800$~GeV, mainly because of the enhanced leptonic branching fractions
of charginos and neutralinos \cite{BCPT}. This feature is common to
the case where $R$-parity is conserved.
 
\item Over most of the region of the $m_0-m_{1/2}$ plane where 
$m_{\tg}$ and $m_{\tq}<1000$ GeV, there are observable signals 
in {\it all} the multilepton channels. 
In our previous analysis, with a stable LSP, the maximal reach was attained in
the $1\ell$ channel. In the current case, the reach in the various leptonic
channels is qualitatively similar. It is reasonable to
suppose that as the charge multiplicity of the leptonic channel increases, 
there are typically more accompanying neutrinos, 
so that the signal events can pass the $\eslt$ requirement
more easily. Nevertheless, a comparison with Fig.~18 of Ref.\cite{BCPT}
shows that the reach in $m_{1/2}$
via the multilepton channels is reduced by $\sim 100-150$~GeV relative
to the $R$-parity conserving case.

\item The regions where the leptonic signals occur extend well beyond
where the $\tz_2$ ``spoiler
modes'' $\tz_2 \to \tz_1 Z$, $\tz_2 \to \tz_1 H_{\ell}$
turn on. This indicates that a large fraction of the events come from
chargino and $W$-boson sources which have accompanying neutrinos. 

\item In the region of parameter space where the OS dilepton signal
has significant contributions from the leptonic decay of $\tz_2$ produced
in a SUSY event cascade, we expect a large number of $e\bar{e}$ and
$\mu \bar{\mu}$ pairs compared to $e\bar{\mu}$ or $\mu \bar{e}$ pairs; in contrast,
for cascade decays dominantly involving leptonic decays of charginos or 
top quarks, we expect essentially the same number of events in each of
the four dilepton flavour channels.
The dilepton flavour asymmetry 
$A_F=\frac{N_{e\bar{e}}+N_{\mu \bar{\mu}}-
N_{e \bar{\mu}}-N_{\mu \bar{e}}}{N_{e\bar{e}}+N_{\mu \bar{\mu}}+N_{e
\bar{\mu}}+N_{\mu \bar{e}}}$ in OS dilepton events is thus a good
indicator\cite{BDKNT} of neutralino production in cascade decay chains.
In Ref.\cite{BCPT} we had shown that for the case of the $R$-parity conserving
scenario, $A_F$ is large and observable over regions of parameter space
where $\tz_2$ has a significant leptonic branching fraction. In contrast,
for the
$R$-violating scenario that we have been studying here, we find that $A_F$ is
small even in the small $m_0$ region where the leptonic decays of
$\tz_2$ are enhanced. We attribute this to the fact that these events
generally do not contain neutrinos, and so, fail
to satisfy the $\eslt$ requirement. 

\item  It would be of interest to investigate 
whether it is possible to 
explicitly reconstruct the $3j$ mass bump for LSP decays,
especially in the SS and $3\ell$ channels which have rather
small backgrounds from Standard Model sources.
This may be difficult due to additional jets from the rest of the event, 
jet mergers, and 
the formidable combinatorial background from the large jet multiplicity
(which would be the distinguishing characteristic of this scenario) in every
SUSY event.

\end{itemize}

Before proceeding further, a technical remark about the simulation leading
to the results in Fig.~1 is worth noting. For the present case, the efficiency
for the signal events to pass the cuts (especially the $\eslt$ cut) is
considerably smaller than in the $R$-parity conserving case studied in
Ref.\cite{BCPT}. As a result, especially for the small $m_{1/2}$ points,
we sometimes have just about 5-10 events that pass all the cuts in our
simulation, so that the signal cross sections have statistical errors as
high as 30-50\%. In contrast, for $m_{1/2} \geq 300$~GeV, we typically
have several tens (or even hundreds) of events in our simulation, at least
for $E_T^c=100$~GeV.
We have checked, however, that except for the row of
points with $m_{1/2}=100$~GeV, there are at least nine signal events in
at least one of the leptonic channels, so that the detectability of the
signal is assured. {\it This is not, however, the case for $m_{1/2}=100$~GeV
for which our simulation does not provide a definite answer because we have
been unable to obtain sufficient statistics. This is not merely academic,
since it leaves open the possibility that there could be a window where 
SUSY might escape detection at the Tevatron as well as the LHC.} 

Finally, we turn to the multijet $+\eslt$ signal for which we have 
used an isolated
lepton veto. The efficiency for this signal is indeed quite small, and it
is difficult to obtain a sufficiently large event sample to be able to
present a contour as in Fig.~1. We have, therefore,
chosen to present our results in
terms of number of events in our simulation that pass our cuts for those
points where the signal is observable according to our criteria. We leave it
to the reader to judge the quality of the simulation and to make a personal
assessment of the reach in the multijet $+\eslt$ channel
from this data. Points where we obtain an observable signal
and for which we have at least nine events in our simulation are denoted by
black squares. We consider points for which there are less than four events
in our simulation as unreliable, and denote these by open squares. Gray squares
denote points  in between these two; here we leave it to the reader to assess
the observability. Finally, the x's denote the regions of the plane
where the signal falls below the observable level. The results of our
simulation are shown in Fig.~2 for the same four cases ({\it a})-({\it d})
as in Fig.~1. Although it may be difficult to assess the exact reach
in this channel, it is clear that as anticipated,
the reach is significantly smaller than in the leptonic channels.
Moreover, even if we assume that there is a reach even for the points denoted
by the grey squares, we see that there are significant ranges of parameters
favoured by fine-tuning considerations 
where there will be no signal in the $\eslt$ channel.

Before closing, we mention that the scenario we have devised may not
be the absolute worst case scenario from point of view of detection
of SUSY signals at the LHC. If baryon number violating couplings are
large these could
modify sparticle decay patterns: In particular, $R$-parity violating
decays of gluinos $\tg \to q\tq^* \to qqq$, or squarks, or of even charginos
and neutralinos,
which do not lead to leptons in the final state could reduce the branching
fractions for the usual decay chains and further reduce the leptonic cross
sections. Any such analysis would be extremely model-dependent and is 
beyond the scope of this paper \cite{BTW}.

To summarize, we have shown that if gluinos and squarks are
lighter than $\sim 1$~TeV, initial searches for
sparticles at the LHC should be able
to detect gluino and squark signals even if $R$-parity is
not conserved, and the LSP decays purely
hadronically via baryon number violating operators.
A clear signal should be seen above Standard Model backgrounds in
the $1\ell$ as well as in several of the multilepton channels. The
signal in the $\eslt$ channel will be significantly reduced relative
to expectations within an $R$-parity conserving framework, and may even be
unobservable if such a scenario is realized in nature. In this case, a
re-analysis of the LHC data with cuts optimized for 
$R$-violating decays could further
improve the signal to background ratio. The main message of this analysis,
however, is that it is unlikely that supersymmetry can remain hidden at
the LHC if squarks and gluinos are lighter than 1~TeV, the preferred mass
range if SUSY is to stabilize the electroweak symmetry breaking sector.

\acknowledgments

This research was supported in part by the U.~S. Department of Energy
under grant numbers DE-FG-05-87ER40319,
DE-FG-03-94ER40833, and DE-FG-02-91ER40685. 

%

%
\newpage

%
\newpage
\begin{figure}
\caption[]{LHC reach contours in the $m_0\ vs.\ m_{1/2}$ plane for various
multi-jet plus multi-lepton plus $\eslt$ signals in the minimal SUGRA model,
but with hadronic $\tz_1$ decays. We show plots for {\it a}) $\tan\beta =2$,
$\mu <0$, {\it b}) $\tan\beta =2$, $\mu >0$, {\it c}) $\tan\beta =10$, $\mu <0$
and {\it d}) $\tan\beta =10$, $\mu >0$. For all frames, we take $A_0=0$ and
$m_t =170$ GeV. 
The $1\ell$ signal is denoted by solid contours, $OS$ dileptons by large dashes,
$SS$ dileptons by small dashes, and the $3\ell$ signal by  dot-dashed contours. 
As discussed in the text, our simulation does not allow
us to reach a definitive conclusion regarding the observability of the signal
for the values of $m_{1/2}$ close to 100~GeV. }
\end{figure}

\begin{figure}
\caption[]{Regions of the $m_0\ vs.\ m_{1/2}$ plane where the multijet$+\eslt$
signal should be observable when $R$-parity conservation is violated and
$\tz_1$ decays hadronically. The frames are the same 
as in Fig. 1. The various squares represent points where the signal is 
observable
according to the criteria stated in the text,
and x's represent points where the signal is invisible. The statistical 
significance of each point is indicated: black squares are where at least 9 
events pass all the simulation cuts, gray squares have 4-8 events passing, 
and open
squares have just 1-3 events passing, and so have the largest statistical 
uncertainty.}
\end{figure}


\begin{references}
\bibitem{FN1} It has been argued that linear $e^+e^-$ colliders operating at a
$\sqrt{s} \sim 1000$~GeV have a
reach comparable to the LHC. This is under the usual assumption that
uncoloured sparticles are significantly lighter than their coloured cousins.
The situation would be very different if the coloured sparticles are the
light ones.
\bibitem{FN3} 
Of course, if this lifetime is sufficiently long,
it may be possible to see the displaced vertex from LSP decay and make 
the signal detection easier.
\end{references}
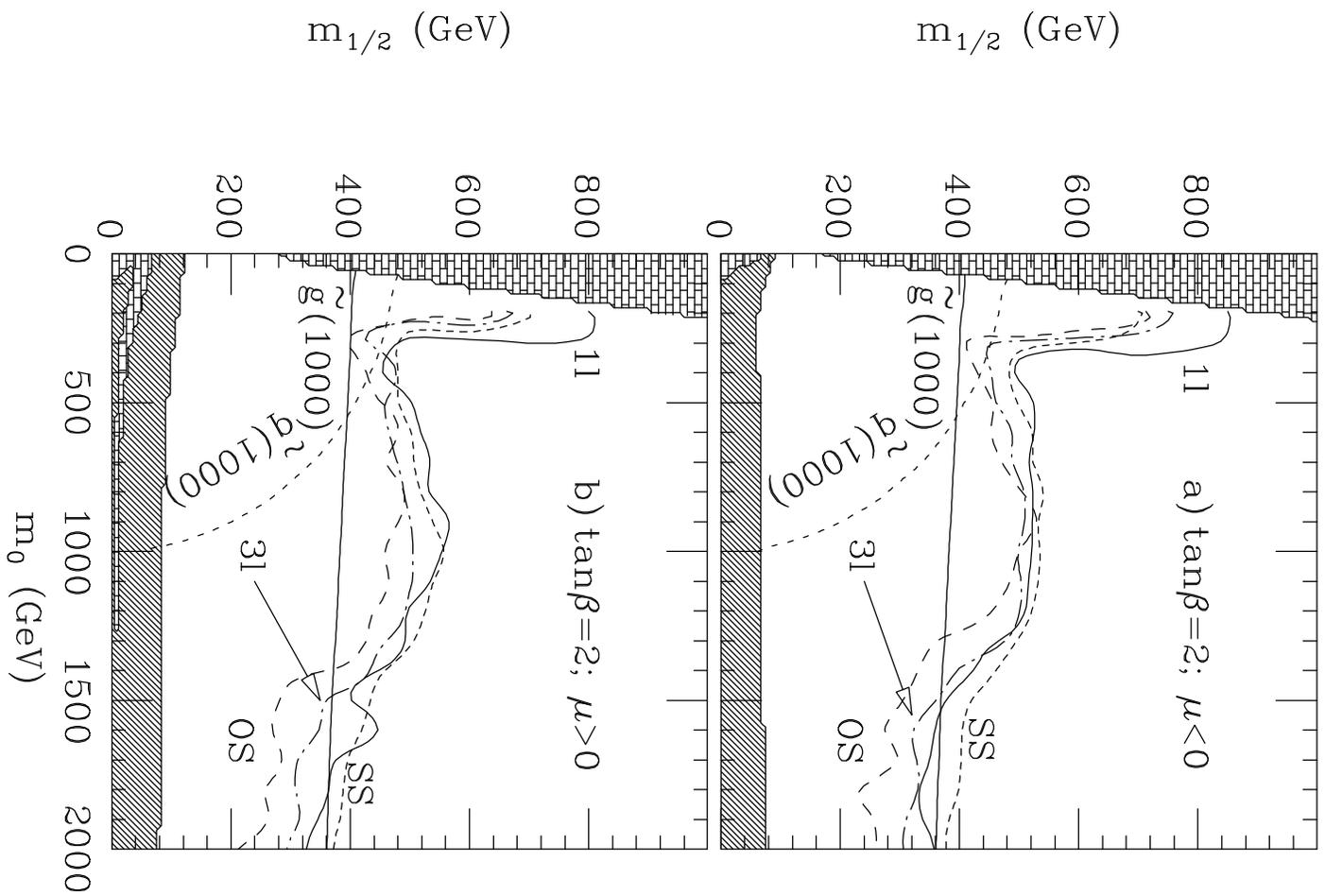
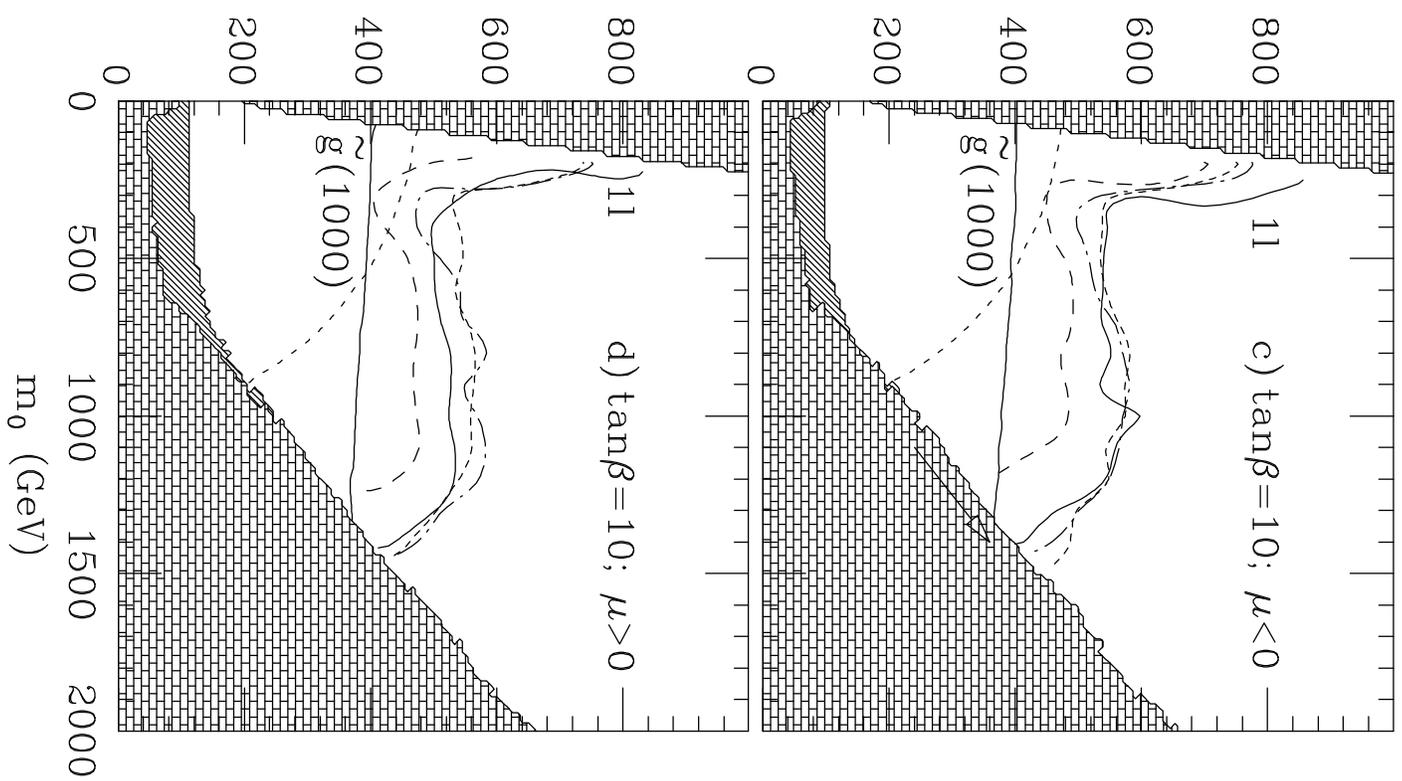
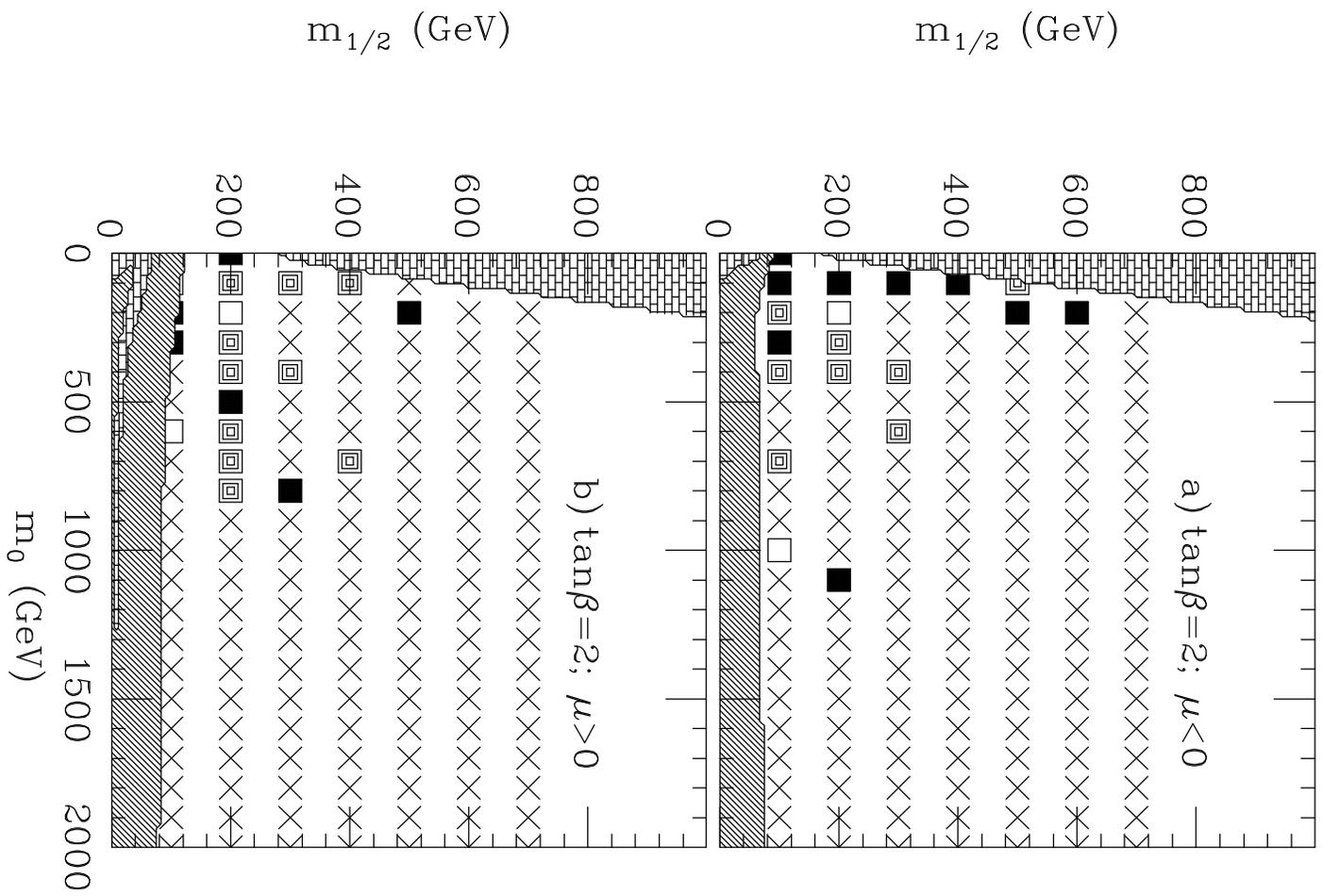
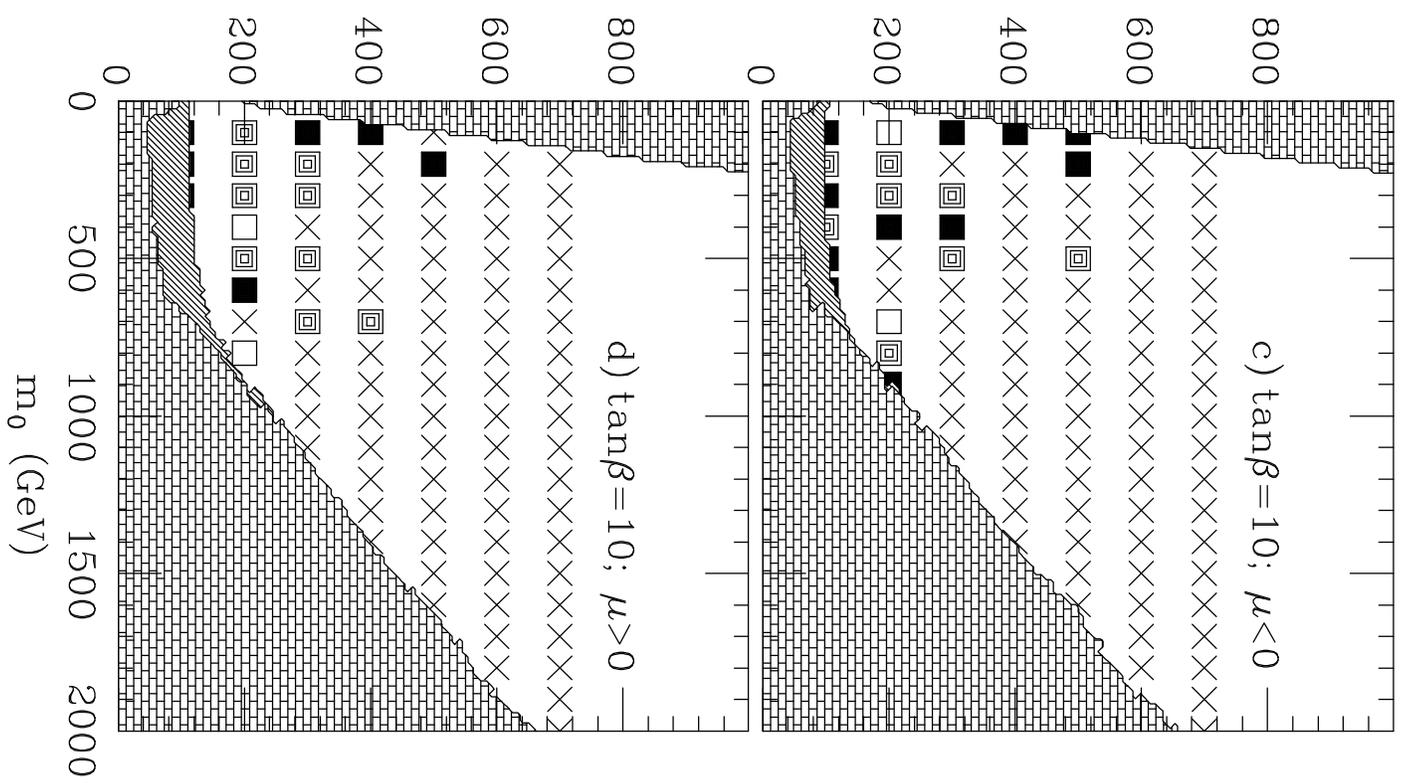
\end{document}